\documentclass[onecollarge,natbib]{svjour2}
\bibpunct{[}{]}{;}{n}{}{,} 
\smartqed  
\usepackage{graphicx}
\usepackage{mathptmx}      
\journalname{Few-Body Systems (EFB22)}
\begin{document}

\title{Implicit vs Explicit renormalization of the $NN$ force
}
\subtitle{An S-wave toy model}


\author{   E. Ruiz Arriola      \and S. Szpigel
         \and  V. S. Tim\'oteo   
}


\institute{Varese Salvador Tim\'oteo \at
              Grupo de \'Optica e Modelagem Num\'erica - GOMNI, Faculdade de Tecnologia - FT, \\ 
               Universidade Estadual de Campinas - UNICAMP,
                     13484-332 Limeira, SP, Brasil \\
              \email{varese@ft.unicamp.br}           
                          \and
           S\'ergio Szpigel \at
              Faculdade de Computa\c c\~ao e Inform\'atica, Universidade Presbiteriana Mackenzie,  \\    01302-907 S\~ao Paulo, SP, Brasil \\
              \email{szpigel@mackenzie.com.br}
              \and
           Enrique Ruiz Arriola \at
              Departamento de F\'\i sica At\'omica, Molecular y Nuclear and Instituto Carlos I \\de F\'\i sica Te\'orica y Computacional. Universidad de Granada, E-18071 Granada, Spain \\
              \email{earriola@ugr.es}
}

\date{Presented by V.S.T. at the 22nd European Conference on Few-Body Problems in Physics (EFB22), 
Krakow, Poland, Sep 2013.}

\maketitle

\begin{abstract}
We use an S-wave toy model for the two-nucleon system to show that the
implicit renormalization of a contact theory matches the explicit renormalization
through a flow equation which integrates out the high momentum components. 
By fitting the low-momentum interaction with a new contact theory, we show
that the running of the contact strengths in both original and fitted contact theories 
match over a wide cutoff range.
\end{abstract}

\section{Introduction}
\label{intro}

The pioneering works by Epelbaum, Gl\"ockle and Meissner have opened 
a new approach for effective field theories based on low-momentum interactions obtained from
unitary transformations \cite{Elow1,Elow2,Elow3}. One of the schemes that has emerged is the 
Similarity Renormalization Group (SRG), which  has been intensively applied to $NN$ interactions 
to produce soft momentum-space potentials that are more suitable to be used as input for nuclear 
structure calculations \cite{Ohio}.   

The similarity renormalization group has also been studied from the point of view of long distance 
symmetries and it was found that the evolved interactions exhibit the Wigner symmetry at a particular 
renormalization scale \cite{SRGsym}. Also, the SRG was applied to the leading order $NN$ interaction 
within the subtractive renormalization framework \cite{SRG-SKM}.  

In this work we want to compare the implicit renormalization of a contact theory at NLO to the 
explicit renormalization of a toy model. The implicit renormalization of the contact theory is performed by constraining the strength of the contact interactions with low-energy parameters like the two-body scattering length and effective range. The explicit integration of the high momentum components can be achieved with the evolution of the $NN$ potential with the similarity renormalization group flow equation with the Block-Diagonal (BD) generator. 
 
\section{SRG with the Block-Diagonal generator}
\label{sub}

The SRG approach is based on a non-perturbative flow equation that governs the evolution of a hamiltonian 
$H=T_{\rm rel}+V$ through an unitary transformation. In operator space, the flow equation is written as 
\begin{eqnarray}
\frac{dH_s}{ds} = \left[  \eta_s, H_s      \right]   \; 
\end{eqnarray}
where $s$ is the flow parameter that ranges from $0$ to $\infty$ and 
\begin{equation}
\eta_s = \left[  G_s, H_s      \right]   \;   
\end{equation}
is the generator of the unitary transformations.

Here we use the Block-Diagonal (BD) generator \cite{BD} as the unitary version of the $V_{\rm{low~k}}$ to 
explicitly integrate out the high momentum components of the Toy Model presented recently in 
Ref. \cite{ImpExp}. The BD generator is given by 
\begin{equation}
G_s = H_s^{BD} \equiv 
\left(
\begin{array}{cc}
  PH_sP   & 0   \\
 0   & QH_sQ    
\end{array}
\right) \;\; ,
\end{equation}
where $P$ and $Q=1-P$ are the projection operators for the low-momentum and the high-momentum spaces respectively. The flow parameter $s$ can be related to a momentum scale called the similarity cutoff,  
$\lambda = s^{-1/4}$, so that when $s$ is evolved towards infinity, the similarity cutoff $\lambda$ approaches zero.

In the infra-red limit of the similarity cutoff, $\lambda \rightarrow 0~(s\rightarrow\infty)$, 
the evolved interaction acquires a block-diagonal form
\begin{equation}
\lim_{\lambda \to 0} V_\lambda = 
\left(
\begin{array}{cc}
  V_{\rm{low~k}}   & 0   \\
 0   & V_{\rm{high~k}}   
\end{array}
\right) \;\; ,
\end{equation}
and the phase shifts produced by the evolved interaction are also separated according to the 
low and high momentum components:
\begin{equation}
\lim_{\lambda \to 0} \delta_\lambda (p) = 
\delta(p)_{\rm{low~k}}  + \delta(p)_{\rm{high~k}}   \;\; .
\end{equation}

The difference between the block-diagonal SRG and the standard $V_{\rm{low~k}}$ approach is that 
the SRG transformation is unitary and the phase shifts corresponding to the high-momentum components, 
$\delta(p)_{\rm{high~k}}$, are also preserved. This is not the case in the $V_{\rm{low~k}}$ prescription, where 
only the low-momentum part of the phase shifts, $\delta(p)_{\rm{low~k}}$ is preserved.

\section{Numerical Results and Discussion}
\label{res}

The question 
we want to answer is the following: how close the low-momentum interaction obtained by integrating out
the high-momentum components is to a contact theory with its low-energy constants constrained by the 
two-body scattering length and effective range?

In order to answer this question, we proceed as follows. First, we use a pionless theory at NLO regulated 
with a smooth gaussian function,  ${\cal R}(\Lambda) = \rm{exp}[-p^{2n}/\Lambda^{2n}]$, 
and adjust the low-energy constants $C^{(2)}_0$ and $C^{(2)}_2$ to obtain the correct scattering length 
and effective range in the $^1S_0$ and $^3S_1$ channels for several values of the cutoff scale $\Lambda$. 

\begin{figure}[h]
\centering
  \includegraphics[width=0.85\textwidth]{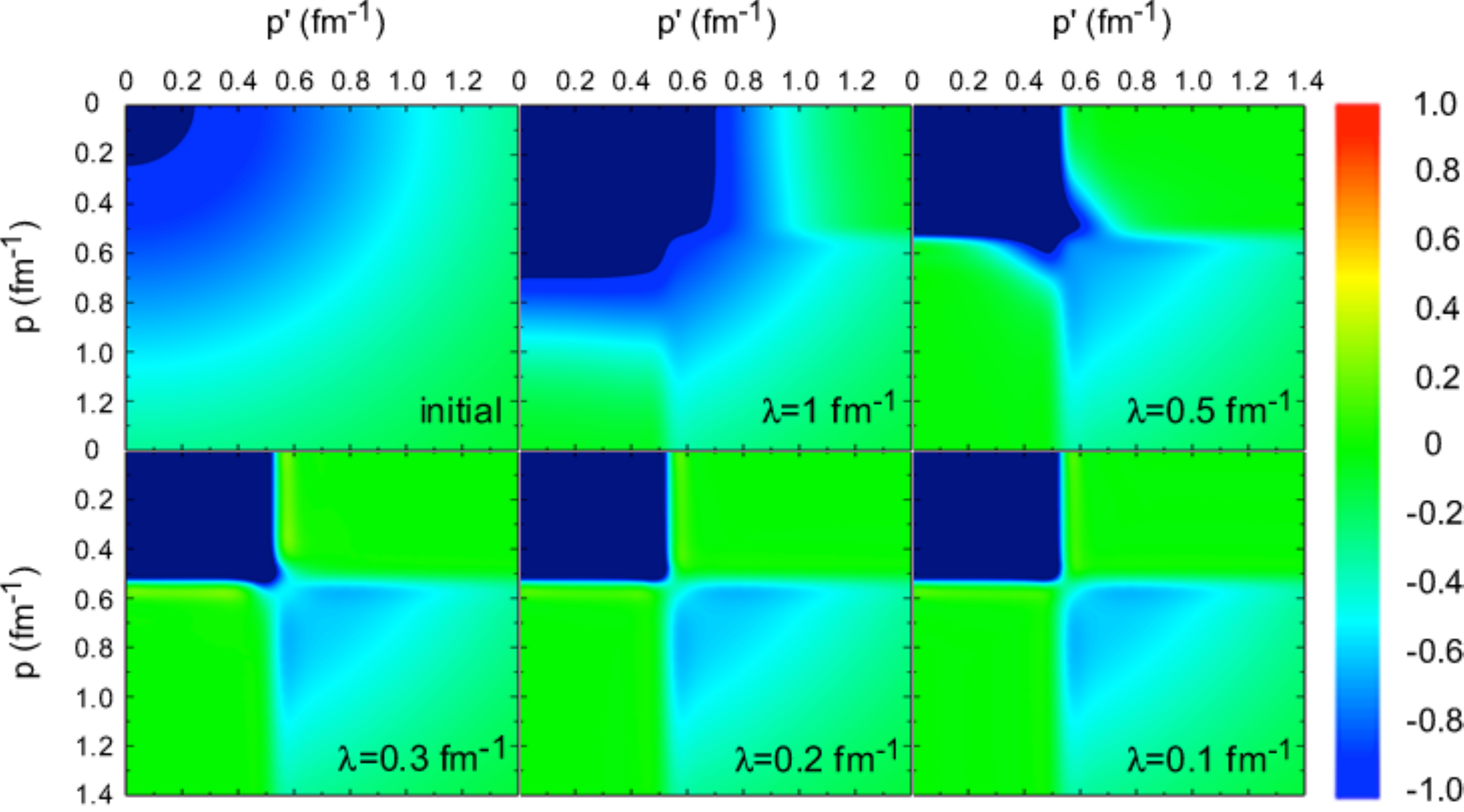}
\caption{Evolution of the toy potential in the $^1S_0$ channel with the block-diagonal generator 
towards the infrared limit of the similarity cutoff $\lambda$. At $\lambda = 0.1~ \rm{fm}^{-1}$, the 
$V_{\rm{low~k}}$ and the $V_{\rm{high~k}}$ parts are completely separated.}
\label{BDev1S0}       
\end{figure}

\begin{figure}[h]
\centering
  \includegraphics[width=0.85\textwidth]{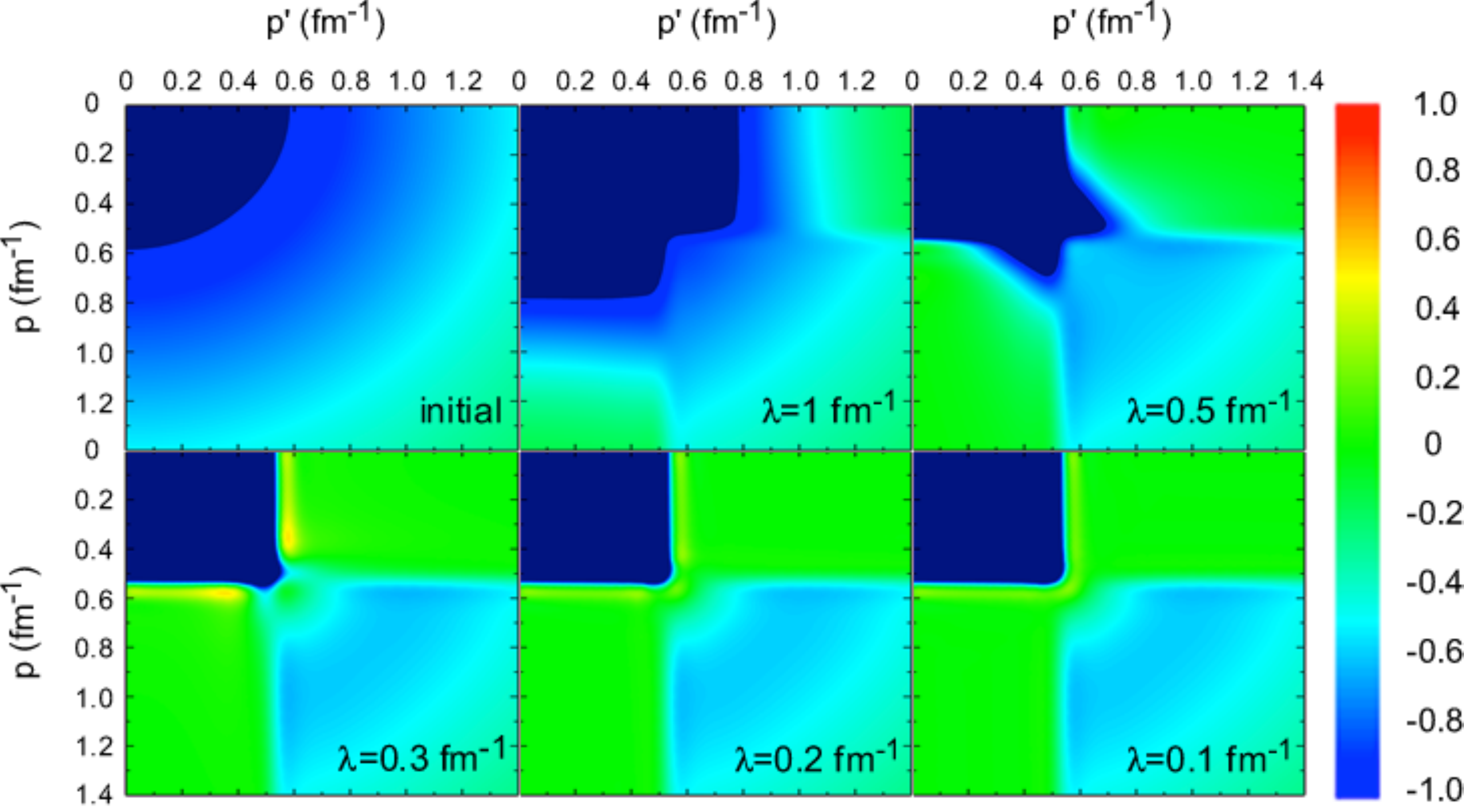}
\caption{Evolution of the toy potential in the $^3S_1$ channel with the block-diagonal generator 
towards the infrared limit of the similarity cutoff $\lambda$. At $\lambda = 0.1~ \rm{fm}^{-1}$, the 
$V_{\rm{low~k}}$ and the $V_{\rm{high~k}}$ parts are completely separated.}
\label{BDev3S1}       
\end{figure}

Then we perform the SRG evolution of our $S$-wave toy model with the block-diagonal generator. 
In Figs. \ref{BDev1S0} and \ref{BDev3S1} we show the evolution of our toy potential in the 
$^1S_0$ and $^3S_1$ channels towards the infra-red limit of the similarity cutoff where we observe 
a complete separation of the low-momentum and high-momentum components as the similarity cutoff approaches zero.

Next, we take the low-momentum components of the evolved interaction up to $\lambda = 0.1~\rm{fm}^{-1}$ 
and fit their diagonal parts at very low momenta with a contact theory at next-to-leading order to determine 
the new low-energy constants $C^{(2)}_0$ and $C^{(2)}_2$. Finally, we compare the running of the 
low-energy constants in both the implicit renormalization and explicit renormalization approaches. 

\begin{figure}[h]
\centering
  \includegraphics[width=0.85\textwidth]{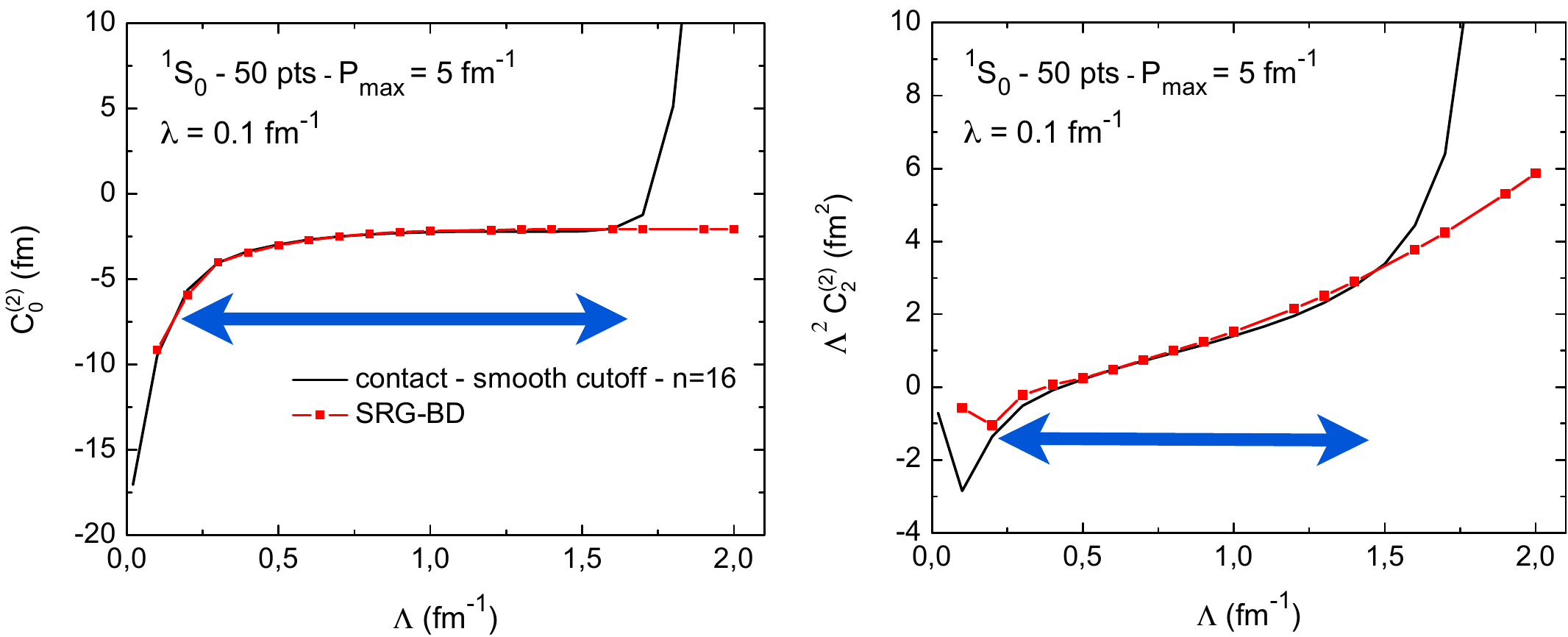}
\caption{Running of the contact strengths in the $^1S_0$ channel with the $V_{\rm{low~k}}$ cutoff $\Lambda$ in the original (black) and in the fitted (red) contact theories at the infrared limit $\lambda = 0.1~ \rm{fm}^{-1}$. 
The blue arrows indicate the range in which they match.}
\label{Run1S0}       
\end{figure}

\begin{figure}[h]
\centering
  \includegraphics[width=0.85\textwidth]{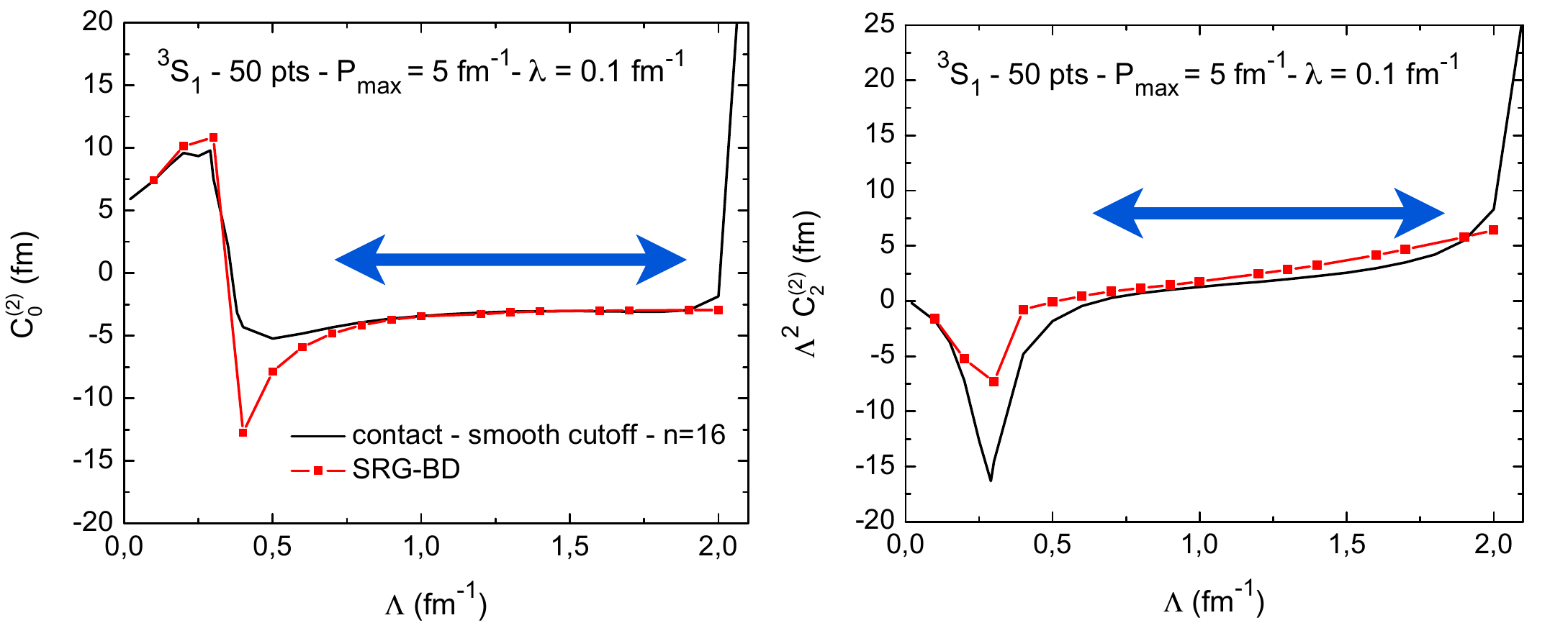}
\caption{Running of the contact strengths in the $^3S_1$ channel with the $V_{\rm{low~k}}$ cutoff $\Lambda$ in the original (black) and in the fitted (red) contact theories at the infrared limit $\lambda = 0.1~ \rm{fm}^{-1}$. 
The blue arrows indicate the range in which they match.}
\label{Run3S1}       
\end{figure}

The result of this procedure can be seen in Figs. \ref{Run1S0} and \ref{Run3S1}, where we display the 
running of the contact strengths with the cutoff scale $\Lambda$ in the $^1S_0$ and $^3S_1$ channels. 
Considering the simplicity of our toy model for the $NN$ interaction, the match of the implicit and explicit 
runnings over a wide cutoff range is impressive.

%

%
%

\section{Concluding Remarks}
\label{conc}

We made a direct comparison of the implicit and explicit renormalization approaches by analyzing 
the running of the low-energy constants obtained in both ways and verified that they match over a
wide cutoff range. In order to overcome the numerical difficulties that appear when the SRG evolution 
is pushed towards the infra-red region of the similarity cutoff, we use a toy model which gives a good 
description the $S$-waves with few discretization points. With the simple model we are able to reduce 
the numerical effort in the solution of the SRG flow equation. 

\begin{acknowledgements}
E.R.A. would like to thank the Spanish DGI (grant FIS2011 - 24149) and Junta de Andalucia 
(grant FQM225). S.S. is partially supported by FAPESP and V.S.T. thanks FAEPEX, FAPESP and CNPq 
for financial support. Computational power provided by FAPESP grant 2011/18211-2.
\end{acknowledgements}

\end{document}